\documentclass[conference]{IEEEtran}
\IEEEoverridecommandlockouts
\usepackage{amsmath,amssymb,amsfonts}
\usepackage{algorithmic}
\usepackage{graphicx}
\usepackage{textcomp}
\usepackage{xcolor}
\def\BibTeX{{\rm B\kern-.05em{\sc i\kern-.025em b}\kern-.08em
T\kern-.1667em\lower.7ex\hbox{E}\kern-.125emX}}

\usepackage{tabularx, multirow}
\usepackage{array}

\graphicspath{{./images/}} 
\DeclareGraphicsExtensions{.pdf,.jpg,.png,.jpeg}

\usepackage[
style=ieee,doi=false,isbn=false,url=false,eprint=false,maxnames=4
]{biblatex}
\AtBeginBibliography{\small}
\begin{document}
	
\title{~\vspace{-12mm}\\{\normalsize Author preprint January 2025}\\
\vspace{-1mm}
The improvement in transmission resilience metrics from reduced outages or faster restoration can be calculated by rerunning historical outage data}

\author{
\IEEEauthorblockN{Arslan Ahmad\hspace{8mm}Ian Dobson}
\IEEEauthorblockA{\textit{Iowa State University} \\
Ames IA USA \\
arslan@iastate.edu, dobson@iastate.edu}
\and
\IEEEauthorblockN{Svetlana Ekisheva\hspace{8mm}Christopher Claypool\hspace{8mm}Mark Lauby}
\IEEEauthorblockA{\textit{North American Electric Reliability Corporation} \\
Atlanta GA USA \\
svetlana.ekisheva@nerc.net, christopher.claypool@nerc.net}
\thanks{AA\,,\,ID gratefully acknowledge support from PSERC, Iowa State University EPRC, and NSF grant 2153163.}
}

\maketitle	

\begin{abstract}
 
Transmission utilities routinely collect detailed outage data, including resilience events in which outages bunch due to weather. The resilience events and associated metrics can readily be extracted from this historical outage data. Improvements such as asset hardening or investments in restoration lead to reduced outages or faster restoration. In this paper, we show how to rerun the historical events including the effects of the reduced outages or faster restorations to measure the resulting improvement in resilience metrics, thus quantifying the benefits of these investments. This is demonstrated with case studies for specific events (a derecho and a hurricane), and all large events or large thunderstorms in the Midwest USA. Instead of predicting future extreme events with models, which is very challenging, rerunning historical events readily quantifies the benefits of resilience investments if these investments had been made in the past. Rerunning historical events is particularly vivid in making the case for resilience investments as it quantifies the benefits for events actually experienced, rather than for uncertain future events.

\end{abstract}


\section{Introduction}

The frequency, severity, and damage of extreme weather events are gradually increasing \cite{NCEI} so it is becoming increasingly important to consider resilience when investing in transmission system hardening and restoration. 
It would be particularly useful to quantify the resilience benefits of proposed upgrades to assets or procedures, so that the resilience benefits could be determined and considered along with all the other factors involved in making system improvements. 
Useful reviews and concepts of power grid resilience include \cite{StankovicPS23,PanteliPS17,MujjuniIA23,NanRESS17}.

While substantial progress has been made in characterizing resilience  with simulation models of weather impacts and 
transmission systems' response and restoration, it remains challenging to quantify resilience and predict the future effect of proposed upgrades with simulations.
These challenges include modeling the complicated processes of resilience from weather conditions, impact on the grid, and operator response, and finally to the restoration processes. This is especially challenging due to the uncertain and heavy-tailed nature of extreme events. The high-impact extreme events are complicated and uncommon but not rare enough to be outliers that will not recur. Indeed, the large transmission system events recur sufficiently often that they exceed the risk of medium-size events \cite{CarrerasPS16}. 

However, there is another promising and practical new approach, driven directly by utility data to quantify resilience and the benefits of resilience upgrades. This approach is to rerun the historical outage events with the system upgrades included. As explained more fully in section \ref{rerun}, the resilience metrics of the actual historical outages can be compared with the improved resilience metrics when the upgrades are included to quantify the resilience benefits. This rerunning history approach avoids modeling errors and is much easier than model-based approaches.

The historical rerun was initially developed for distribution systems \cite{AhmadPS24}, and the main contribution of this paper is to show how it can also be used for transmission system resilience.
In particular, we quantify the benefits that reduced outages or faster restorations would have had in the following North American cases:
\begin{itemize}
    \item Particular extreme events: a derecho and a hurricane
    \item All large events in the Midwest USA over a time period. 
    \item A specific type of large event: thunderstorms in the Midwest USA.
\end{itemize}
These cases enable several different accounts or perspectives of the range of benefits.

We note that other aspects of transmission system resilience can be quantified with utility data:
Outage and restore process and performance curve  metrics are extracted and applied in 
\cite{EkishevaISGT23} and used to model typical restore processes with nonhomogeneous Poisson processes in \cite{DobsonPS24,DobsonPESL23}.
Kelly-Gorham \cite{KellyEPSR20} samples from empirical probability distributions to model and quantify resilience processes.

\section{Extracting metrics from TADS outage data}

The North American Electric Reliability Corporation (NERC) has collected inventory and outage data for transmission elements with voltages above 100 kV since 2015. These data are reported by transmission owners across North America to NERC’s Transmission Availability Data System (TADS) \cite{nercTADS}.
Additional data processing of TADS data is needed to identify large transmission outage events that typically affect transmission equipment owned by several utilities. A grouping algorithm that joins TADS outages that started in quick succession allows us to identify large outage events \cite{EkishevaPESGM21}. Using external weather data sources (e.g., NOAA), we identify weather-related large outage events and classify them by extreme weather type \cite{EkishevaPESGM21}. Assessing system performance in terms of events rather than individual outages is one of the characteristics of resilience analysis. 

For each outage event, we extract outage and restore processes and a performance curve; these processes enable us to calculate several important statistics that quantify transmission system resilience \cite{EkishevaPMAPS22, DobsonPS24}.
The event size (the number of outages in the event), the number of distinct TADS elements affected, total transmission capacity (MVA) and circuit-miles affected, time to first restore, the maximum number of elements out, maximum amount of MVA out, element-day loss and MVA-day loss are the metrics that gauge the magnitude and intensity of the extreme weather as well as grid’s response to it. 
Additionally, the event duration and the restore duration are statistics describing the system’s recovery; however, due to their statistical volatility, other duration metrics are preferable \cite{DobsonPS24}, such as time to substantial restoration level, when either 95\% of transmission outages or 95\% of MVA is restored. 
Beyond statistical practicality, the 95\% restoration level is often a prerequisite for the utilities to re-energize the distribution system to begin serving customer loads \cite{nercSOR23}.

Given the ability to extract transmission system events from standard utility data and calculate the metrics describing the resilience of these events, we now address the challenge of quantifying the benefits of resilience improvements by ``rerunning history". 

\section{Rerunning History} \label{rerun}

In our case studies, we use a ``rerunning history” approach to quantify the benefits of two types of resilience investments: investments made to harden the infrastructure and investments in faster restorations. The historical rerun method gives the improvements in resilience metrics that a proposed resilience investment {\sl would have had} if the investment had been made in the past \cite{AhmadPS24}. 
Since it is driven by real data, this has the advantage of incorporating all the factors affecting resilience over the past period such as weather, vegetation management, human factors, operating procedures, equipment aging, system reconfigurations, and restoration procedures. The historical rerun method has no modeling error from these factors. The historical rerun method does not predict the future, but the methods of predicting the future with simulation models are very complicated, whereas the historical rerun is much simpler. 

Moreover, the historical rerun method has some advantages in making the case for resilience investments to customers and regulators: The benefits that would have applied to the lived experience of stakeholders in the past, both for particular large events and in general, are likely to be more persuasive than the benefits that are simulated for predicted events at some indeterminate time in the future.

\section{Rerun history for specific events}
\looseness=-1
Rerunning historical events can quantify the benefits of investments in hardening and faster restorations for individual outage events. 
This can be applied to large events within the memory of stakeholders to help make the case for resilience investments. 
Especially for recent large events, it can be persuasive to be able to present the benefits of the investment that would have been gained for a specific event that impacted customers' lives.
We select two examples of specific events: the Midwest derecho of August 2020, and Hurricane Ida in 2021. 

For this case study we assume that an investment made to harden the infrastructure would result in a 10\% reduction in the number of outages. Different types of investments can be made to get such 10\% hardening. For example, reinforcing or replacing vulnerable transmission towers and lines, installing high-strength conductors, 
vegetation management, and hardening the substations to minimize damage due to flooding. The type of hardening depends on the landscape of the area where the transmission system is located and the type of threats the transmission system is most exposed to.

To implement the 10\% hardening, we randomly remove 10\% of the outages from the event and recompute the resilience metrics. This step is done for 10\,000 different random removals and the resulting 10\,000 resilience metrics are averaged to find the average resilience metrics obtained with the investment. Then the change in each resilience metric is the average resilience metric with the investment minus the base case resilience metric.

Investments can also be made to improve the restoration process. 
Such investments include up-skilling existing crews and/or hiring more crews to complete the restoration process faster, equipping the crews with more and better resources, improving the post-event restoration planning with automation, and improving the stockpiles of critical components. 
To quantify the benefits of faster restorations, we assume that such an investment would make the restoration 10\% faster. 
We rerun the resilience processing for each event with the same outages but adjust the restoration times to reflect 10\% faster restorations.\footnote{The restore duration, which is the outage restore time minus the time of the first restore in the event, is reduced by 10\%, as long as the new restore time occurs after its outage time; see \cite[sec. 6.3.7]{AhmadPS24}.}

\subsection{Hurricane Ida, August 2021}
\label{Ida}
Hurricane Ida, a Category 4 hurricane, made landfall in Louisiana on August 29, 2021. 
As the hurricane cut across Southeastern Louisiana, it maintained hurricane strength, primarily affecting entities in Louisiana and Mississippi. Hurricane-force winds reaching up to 150~mph were predominately isolated to Louisiana. 
Approximately 1.2 million customers lost power in SERC Reliability Corporation's footprint, including the greater New Orleans area. In New Orleans, all eight transmission lines serving the city were knocked offline. 
Further analysis of Ida is in \cite{EkishevaISGT23}.

\begin{figure}[!t]
    \centering
    \includegraphics[width=0.48\textwidth]{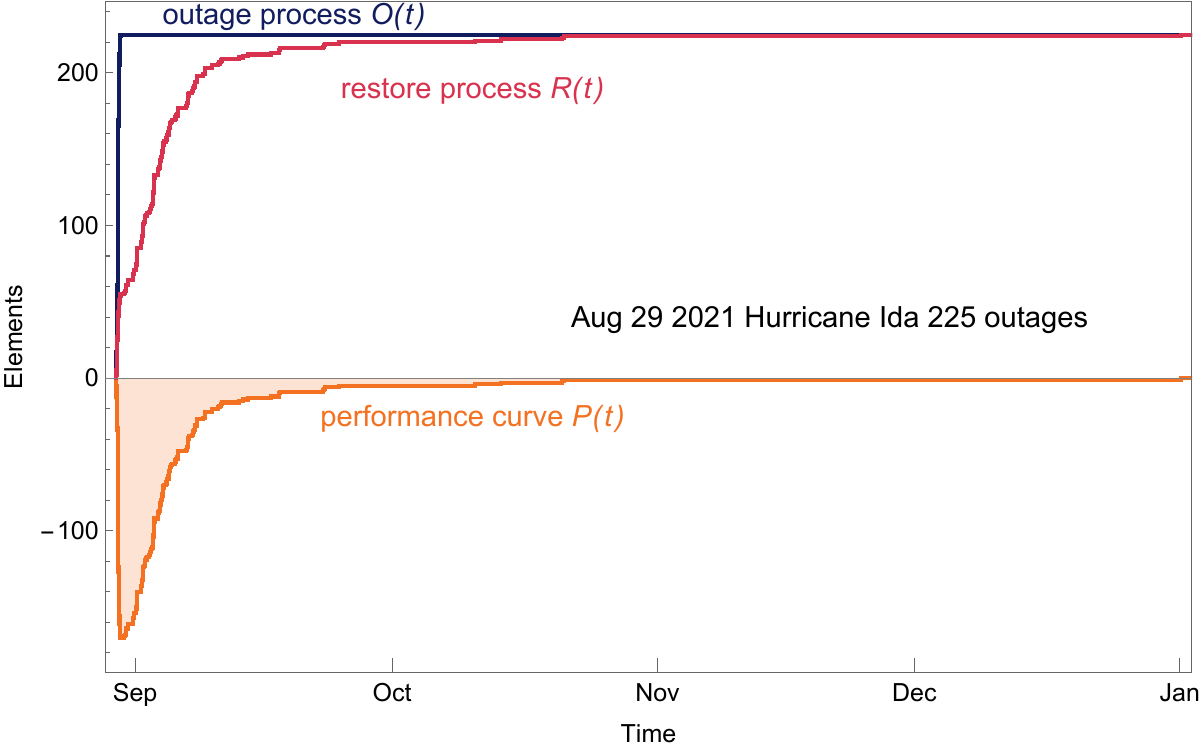}
    \caption{Outage \& restore processes, and performance curve of Hurricane Ida}
    \label{fig:performanceCurveIda}
\end{figure}

The base case resilience metrics of Hurricane Ida, along with the improvement in each resilience metric due to 10\% hardening and 10\% faster restorations, are shown in Table \ref{tab:resultsHurricaneIda}.
The {\sl event size} is the number of outages in the event. 
Since we randomly remove 10\% of the outages\footnote{Since 10\% of 225 is 22.5, 22 outages are randomly removed in 5000 samples and 23 outages are randomly removed in the other 5000 samples to get an overall 10\% reduction in outages.}, the {\sl event size} metric decreases by exactly 10\%.
The {\sl nadir} is the maximum mean number of elements simultaneously outaged during the event, or the negative of the minimum value of the performance curve shown in Fig.~\ref{fig:performanceCurveIda}.
For Hurricane Ida, the {\sl nadir} occurred on 08-30-2021 at 1:40 AM CDT. 
Consider removing an outage from the event due to hardening:
If the outage and its restoration are both before the {\sl nadir}, the {\sl nadir}  will not change. 
Whereas, if an outage is before the nadir and its restoration is after the {\sl nadir}, the {\sl nadir} will decrease by 1.
This means that if the initial outages that take a long time to restore can be identified and mitigated, then a more substantial reduction in the {\sl nadir} can be achieved.

\begin{table}[!t]
    \caption{Resilience Improvement Results for Hurricane Ida}
    \label{tab:resultsHurricaneIda}
    \centering
    \begin{tabular}{c c c c}
        &Base Case&Change with&Change with\\
       Resilience Metric &Value&Hardening&Faster Restorations\\
        \hline
        Event Size          & 225       & -10.0\%       & 0\%      \\
        Nadir (elements)    & 171       & -10.2\%       & 0\%  \\
        $D_{95\%}$ (hours)    & 457.8    & -2.24\%       & -10.0\%  \\
        MVA-days out        & 641506    & -10.2\%       & -10.3\%  \\[2pt]
        \hline
    \end{tabular}
\end{table}

\looseness=-1
The $D_{95\%}$ metric is the time of the 95\% quantile of restoration minus the time of the first restoration as calculated using interpolation in \cite{DobsonPS24}.
$D_{95\%}$ decreases with hardening when the restoration time of a removed outage is  $\ge \!D_{95\%}$; i.e., it falls in the upper 5\% quantile.
Removal of any other outage from the event results in a slight increase in $D_{95\%}$.
In the case of Hurricane Ida,  there are only 12 out of 225 outages with restoration times $\ge \!D_{95\%}$, so these 12 outages are less likely to be randomly removed, that is why overall we see a 2.24\% decrease in $D_{95\%}$ with 10\% hardening in Table~\ref{tab:resultsHurricaneIda}. 


{\sl MVA-days out} is the area under a performance curve drawn with the elements' MVA rating lost on the vertical axis and a time scale of days.
It is also equal to the sum of the products of MVA rating lost and the duration of each outage in the event.
Removal of any outage with a positive duration decreases the {\sl MVA-days out} metric.

\looseness=-1
When one or more outages are removed from an event, the event can occasionally get split into two or more events. We keep track of the split events using the ``Super Event" concept \cite{AhmadPS24}.
The metrics of all the events in a super event are aggregated, with the aggregation method varying with the metric.\footnote{
We aggregate {\sl event size} and {\sl MVA-days out} by adding the corresponding values for each of the split events in the super event. We aggregate {\sl nadir} by taking the maximum {\sl nadir} of all the split events. For $D_{95\%}$, we consider the super event as a single event to calculate $D_{95\%}$.}

The faster restoration results for Hurricane Ida are shown in Table \ref{tab:resultsHurricaneIda}. Faster restorations do not affect the outages so that the {\sl event size} does not change. Faster restorations have little effect or, in this case, no effect on the {\sl nadir}.
10\% faster restorations decreases $D_{95\%}$ by exactly 10\%. 

The change in {\sl MVA-days out} due to faster restorations varies, as it depends on different factors, including the number of momentary outages in the event, whether the outage occurs before or after the first restore in the event, and the duration of each outage. When we decrease the restore duration of an outage by 10\%, it doesn't always reduce the outage duration (and thus {\sl MVA-days out}) of that outage by 10\%.
All the outages that start after the time of the first restore decrease more than 10\% in their {\sl MVA-days out}, whereas all the outages that begin before the time of the first restore decrease less than 10\%. 
Therefore, if more outages start after the time of first restore as compared to the outages that started before the time of first restore, the overall decrease in {\sl MVA-days out} will tend to be greater than 10\%, as in the case of Hurricane Ida.
However, overall the decrease in {\sl MVA-days out} averages to approximately 10\% for a large-size event. 

\subsection{Midwest Derecho, August 2020}
\begin{figure}[!t]
    \centering
    \includegraphics[width=0.48\textwidth]{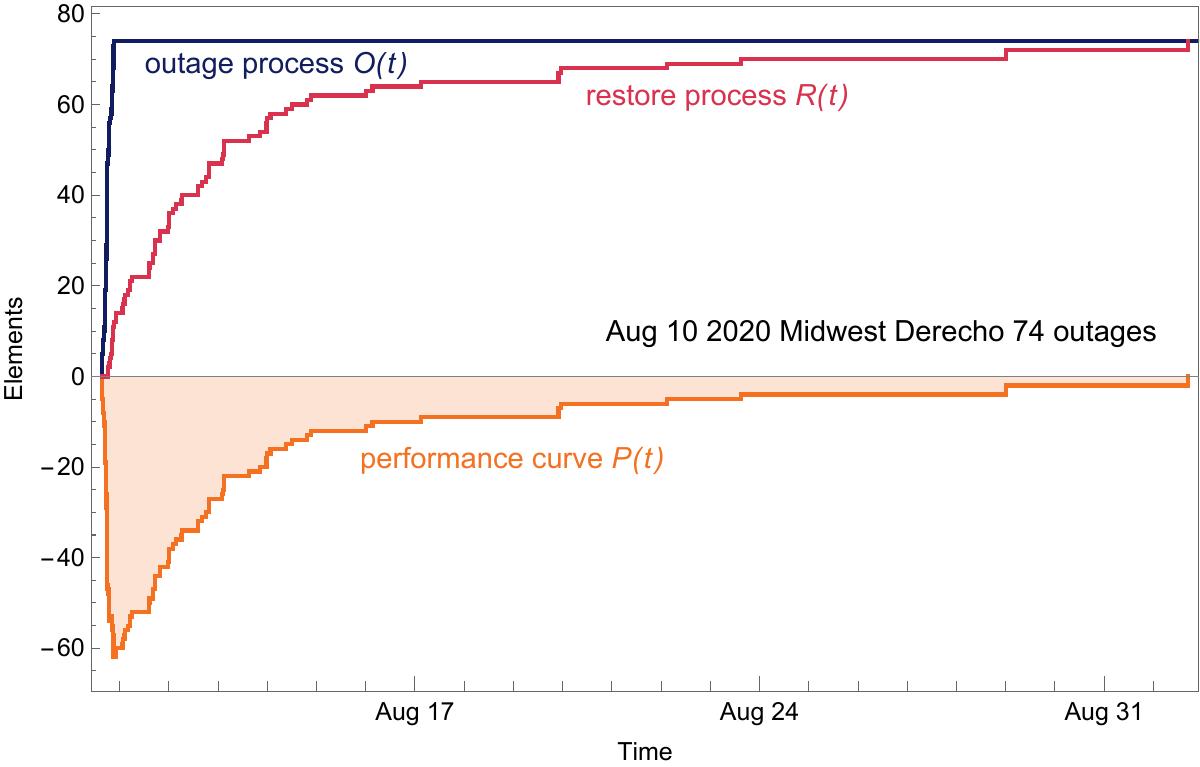}
    \caption{Outage \& restore processes, \& performance curve of Midwest Derecho}
    \label{fig:performanceCurveDerecho}
\end{figure}

On August 10-11, 2020, a powerful derecho swept through the Midwest, bringing hurricane-force winds exceeding 100 mph and causing widespread devastation in multiple states. The storm, spanning over 700 miles, left more than a million customers without power across Iowa, Illinois, and surrounding areas. In Iowa alone, nearly 600,000 people experienced prolonged outages as utility crews faced massive infrastructure damage, including downed lines and debris-blocked access routes. The outage process, restore process, and the performance curve of the derecho are shown in Fig. \ref{fig:performanceCurveDerecho}.


The base case resilience metrics, along with the improvement in each resilience metric due to 10\% hardening and 10\% faster restorations, are shown in Table \ref{tab:resultsMidwestDerecho}. 
The changes in metrics are similar to those in Table \ref{tab:resultsHurricaneIda}, except for $D_{95\%}$. The $D_{95\%}$ decreases more in this case as compared to Hurricane Ida due to the effects explained in section~\ref{Ida}.
In general, larger events with a few outages taking longer times to restore see a smaller decrease in $D_{95\%}$ as a result of hardening as compared to smaller events.

\begin{table}[!t]
    \caption{Resilience Improvement Results for the Midwest Derecho}
    \label{tab:resultsMidwestDerecho}
    \centering
    \begin{tabular}{c c c c}
        &Base Case&Change with&Change with\\
       Resilience Metric &Value&Hardening&Faster Restorations\\
        \hline
        Event Size          & 74       & -10.0\%       & 0\%      \\
        Nadir (Elements)    & 62       & -10.6\%       & 0\%  \\
        $D_{95\%}$ (hours)    & 431.6    & -6.84\%       & -10.0\%  \\
        MVA-days out        & 65693    & -10.1\%       & -9.96\%  \\[2pt]
        \hline
    \end{tabular}
\end{table}

\section{Rerun history for all large events in a region} \label{sec:rerunAllEvents}
Previously, we demonstrated how to calculate the benefits of resilience investments for a single event.
In this section, we extend this method to quantify resilience benefits across all large events within a region of a transmission system.
It is important to note that while this method could theoretically be applied to a vast transmission footprint, such as the entire Eastern Interconnection, doing so will be less useful for a number of reasons.
First, large transmission systems cross multiple jurisdictions, each with distinct regulatory frameworks, operational procedures, and investment priorities. Consequently, analysis at this scale would not accurately capture the varied types of investments and their impacts across such a system.
Second, large transmission systems often span geographically diverse regions with distinct climates and different types of extreme weather events. Applying a uniform resilience analysis across such diverse conditions would produce generalized results that can fail to accurately reflect the localized impacts and benefits of resilience investments in any particular region.
Therefore, to ensure more applicable results, we apply this method to a smaller, more homogeneous region.
Specifically, for our example use case, we select all large (more than 20 outages) weather events in the Midwest\footnote{The  Midwest comprises Illinois, Indiana, Iowa, Kansas, Michigan, Minnesota, Missouri, Nebraska, North Dakota, Ohio, South Dakota, and Wisconsin, as per the U.S. Census Bureau's definition. This differs from the definition of NERC’s Region (Electric Reliability Organization’s Regional Entity).} USA for our analysis.
Details of the large events in the Midwest from 2015 to 2021 are given in Table~\ref{tab:midwestEventDetails}.

\begin{table}[!t]
    \caption{Details of large events in the Midwest Region}
    \label{tab:midwestEventDetails}
    \centering
    \begin{tabular}{c c c c c}
        Event ID&Year&Event Type&Event Size&MVA-days lost\\
        \hline
        1    & 2015    & Winter weather     & 21    & 378             \\
        2    & 2015    & Winter weather     & 62    & 2177   \\
        3    & 2015    & Tornado            & 21    & 11104   \\
        4    & 2015    & Tornado            & 24    & 8224   \\
        5    & 2016    & Thunderstorm       & 21    & 3836   \\
        6    & 2017    & Thunderstorm       & 102   & 217462    \\
        7    & 2018    & Thunderstorm       & 25    & 6743   \\
        8    & 2018    & Tornado            & 42    & 4684  \\
        9    & 2019    & Winter weather     & 20    & 2665   \\
        10   & 2019    & Winter weather     & 49    & 33571   \\
        11   & 2019    & Tornado            & 22    & 17896   \\
        12   & 2020    & Thunderstorm       & 20    & 6398   \\
        13   & 2020    & Thunderstorm       & 58    & 40277   \\
        14   & 2021    & Winter weather     & 21    & 3533   \\
        15   & 2021    & Thunderstorm       & 87    & 57369   \\[2pt]
        \hline
    \end{tabular}
\end{table}
We select all the large events with weather causes in the Midwest region and calculate their base case average resilience metrics.
To represent asset hardening, we randomly remove $10$\% of outages (from all outages in large events), repeat several times, and calculate the average resilience metrics with hardening.
Then, to represent faster restorations, we calculate the average resilience metrics after decreasing the restore duration of all the large events.  
The results with 10\% hardening and 10\% faster restorations are shown in Table \ref{tab:resultsMidwestOverall}. The results of hardening can be compared with the results of faster restorations. The faster restorations do not affect the outages or the nadir. However, it is more effective than hardening in reducing the duration $D_{95\%}$, and roughly equally effective in reducing the MVA-days out.

\begin{table}[!ht]
    \caption{Resilience improvements for 10\% hardening or 10\% faster restorations for large events in the Midwest}
    \label{tab:resultsMidwestOverall}
    \centering
    \begin{tabular}{c c c c}
        &Base Case&Change with&Change with\\
       Resilience Metric &Value&Hardening&Faster Restorations\\
        \hline
        Event Size          & 40       & -10.0\%       & 0\%      \\
        Nadir (Elements)    & 21.4     & -14.9\%       & 0\%  \\
        $D_{95\%}$ (hours)    & 147.7    & 1.39\%       & -10.0\%  \\
        MVA-days out        & 30837    & -10.0\%       & -10.4\%  \\[2pt]
        \hline
    \end{tabular}
\end{table}

\section{Rerun history for a specific type of events}
Transmission system resilience can be enhanced through various investment strategies. Certain investments may improve resilience across a range of large events, while others can be tailored to strengthen the system resilience for specific types of weather events.
In this section, we examine large events caused by thunderstorms in the Midwest and estimate the benefits of 10\% system hardening and a 10\% improvement in restoration by applying the analysis outlined in Section~\ref{sec:rerunAllEvents}. The results are presented in Table~\ref{tab:resultsMidwestThunderstorms}. This method can be similarly applied to events caused by tornadoes or winter weather or hurricanes or other hazards, offering a general way to assess the impact of tailored resilience investments.

\begin{table}[!ht]
    \caption{Resilience improvements for 10\% hardening or 10\% faster restorations for thunderstorms in the Midwest}
    \label{tab:resultsMidwestThunderstorms}
    \centering
    \begin{tabular}{c c c c}
        &Base Case&Change with&Change with\\
       Resilience Metric &Value&Hardening&Faster Restorations\\
        \hline
        Event Size          & 52.2       & -10.0\%       & 0\%      \\
        Nadir (Elements)    & 30.8     & -14.7\%       & 0\%  \\
        $D_{95\%}$ (hours)    & 223.5    & 2.71\%       & -10.0\%  \\
        MVA-days out        & 61503    & -10.0\%       & -10.3\%  \\[2pt]
        \hline
    \end{tabular}
\end{table}

\section{Discussion and Conclusions}

\looseness =-1
When considering transmission system investments that improve resilience by reducing outages or restoring faster, it is useful to be able to quantify the benefits. 
In this paper we show how to rerun historical events while including the effect of reducing outages or restoring faster to calculate the change in resilience metrics. 
Reducing outages not only reduces event size but also affects event duration, often reducing it \cite{nercLL23}, and the historical rerun quantifies these effects. Whereas faster restorations do not affect the outages and the {\sl event size}, and have little or no effect on the {\sl nadir}. Here we analyze the effects of reduced outages and faster restorations separately. However, many upgrade plans would include a combination of reduced outages and faster restorations, which also would be easy to analyze.

The benefits for resilience are computed for specific single events, all events in a region, and for a single type of events in a region. 
This enables flexibility in making a case for investment in resilience. 
The benefits for a single extreme event that the stakeholders can recall from their own experience can be more tangible and persuasive. Or the accumulated benefits over all the recorded large events in a region may be of interest. 

A larger region has more large events and thus more data, but is less representative of a subregion. 
We consider a relatively large region, the Midwest USA, that has some features of geography and extreme weather threats in common to calculate the benefits over the region. 
However, smaller regions such as states could be considered, especially if the benefits needed to be quantified for the state stakeholders. 
While some investments, such as stronger towers or lighter conductors, would increase resilience for a range of threats to resilience, other investments, such as better lightning protection, would target more specific resilience threats.
Therefore, we also show how to calculate the resilience benefits for a particular type of events, such as thunderstorms.

In the case studies, we express the hardening in terms of a percent reduction in outages and a percent increase in the rate of restorations. 
In this paper, we mention a few of the many ways that a utility could achieve these improvements, but do not describe any of the detailed engineering needed to implement the percentage improvements. 
This detailed engineering is important, but must be done in the context of specific improvements for specific grids, taking into account their particular vulnerabilities. 
Utilities already have considerable expertise in this detailed engineering in proposing upgrades to assets or procedures and estimating their impacts on outages or restoration. 
This paper shows a straightforward way to take these estimates and better quantify the resilience benefits of these proposals with metrics by rerunning history. Moreover, the improvements in the metrics can assist in communicating the resilience benefits of proposals to stakeholders.









\printbibliography 

\end{document}